\documentclass[preprints,article,accept,moreauthors,pdftex]{Definitions/mdpi}

\firstpage{1} 
\pubvolume{xx}
\issuenum{1}
\articlenumber{5}
\pubyear{2019}
\copyrightyear{2019}
\history{}

\hypersetup{colorlinks=true, citecolor=blue, urlcolor=blue, linkcolor=blue}
\usepackage[english]{babel}
\usepackage[utf8]{inputenc}
\usepackage{graphicx}
\usepackage[caption=false]{subfig}
\usepackage{amsmath}
\usepackage{amsfonts}
\usepackage{amssymb}
\usepackage{doi}
\usepackage{color,soul}
\setulcolor{red}

\usepackage{xcolor}

\Title{Multiple current reversals using superimposed driven lattices}

\Author{Aritra K. Mukhopadhyay $^{1}$*\orcidA{} and Peter Schmelcher $^{1,2}$*}

\AuthorNames{Aritra K. Mukhopadhyay and Peter Schmelcher}

\address{%
$^{1}$ \quad Zentrum f\"ur Optische Quantentechnologien, Fachbereich Physik, Universit\"at Hamburg, Luruper Chaussee 149, 22761 Hamburg, Germany.\\
$^{2}$ \quad The Hamburg Centre for Ultrafast Imaging, Universit\"at Hamburg, Luruper Chaussee 149, 22761 Hamburg, Germany.}

\corres{Correspondence: Aritra.Mukhopadhyay@physnet.uni-hamburg.de (A.K.M.); Peter.Schmelcher@physnet.uni-hamburg.de (P.S.)}

\abstract{We demonstrate that directed transport of particles in a two dimensional driven lattice can be dynamically reversed multiple times by superimposing additional spatially localized lattices on top of a background lattice. The timescales of such current reversals can be flexibly controlled by adjusting the spatial locations of the superimposed lattices. The key principle behind the current reversals is the conversion of the particle dynamics from chaotic to ballistic, which allow the particles to explore regions of the underlying phase space which are inaccessible otherwise. Our results can be experimentally realized using cold atoms in driven optical lattices and allow for the control of transport of atomic ensembles in such setups.}


\begin{document}


\section{Introduction}
Originally conceived as a proof of principle behind the working of biological motors \cite{AitHaddou_CBB_Brownian_Ratchet_2003,Astumian_PT_Brownian_Motors_2002,Astumian_S_Thermodynamics_Kinetics_1997,Julicher_RMP_Modeling_Molecular_1997}, the phenomenon of `ratchet' transport of particles, i.e. the emergence of unidirectional particle transport in an unbiased non-equilibrium environment, has gained widespread applications across various disciplines \cite{Astumian_PT_Brownian_Motors_2002,Bartussek_EL_Periodically_Rocked_1994,Cubero__Brownian_Ratchets_2016,Denisov_PR_Tunable_Transport_2014,Faucheux_PRL_Optical_Thermal_1995,Hanggi_AP_Brownian_Motors_2005,Hanggi_RMP_Artificial_Brownian_2009,Magnasco_PRL_Forced_Thermal_1993,Prost_PRL_Asymmetric_Pumping_1994,Reichhardt_ARCMP_Ratchet_Effects_2017,Renzoni__Driven_Ratchets_2009}. The necessary ingredients required for such a rectification of random particle motion into directed transport has been shown to be non-equilibrium, non-linearity and the breaking of certain spatio-temporal symmetries \cite{Denisov_PR_Tunable_Transport_2014,Flach_PRL_Directed_Current_2000,Schanz_PRE_Directed_Chaotic_2005}. Since then the ratchet effect has found numerous applications including particle separation based on physical properties \cite{Matthias_N_Asymmetric_Pores_2003,Mukhopadhyay_PRL_Simultaneous_Control_2018,Wambaugh_PRE_Ratchetinduced_Segregation_2002}, design of efficient velocity filters \cite{Petri_EL_Formation_Density_2011,Wulf_PRE_Analysis_Interface_2012}, transportation of fluxons in Josephson junctions arrays \cite{Falo_EL_Ratchet_Potential_1999,Zolotaryuk_PRE_Asymmetric_Ac_2012}, unidirectional motion of active matter \cite{Ai_SR_Ratchet_Transport_2016,Reichhardt_ARCMP_Ratchet_Effects_2017}, voltage rectification in superconducting quantum interference devices (SQUID) \cite{Spiechowicz_CIJNS_SQUID_Ratchet_2019,Spiechowicz_NJP_Efficiency_SQUID_2015,Zapata_PRL_Voltage_Rectification_1996} and enhancement of photocurrents in quantum wells \cite{Faltermeier_PRB_Magnetic_Quantum_2017}.

Due to novel experimental progress in atom trapping techniques, directed transport of atomic ensembles has been realized in ac-driven optical lattices \cite{Lebedev_PRA_Twodimensional_Rocking_2009,Schiavoni_PRL_Phase_Control_2003} both in the ultracold quantum regime \cite{Salger_S_Directed_Transport_2009} and at micro kelvin temperatures where a classical dynamics approach successfully describes the experiments \cite{Brown_PRA_Ratchet_Effect_2008,Renzoni__Driven_Ratchets_2009}. Apart from the vast majority of ratchet based setups in one spatial dimension (1D) \cite{Denisov_PR_Tunable_Transport_2014,Flach_PRL_Directed_Current_2000,Liebchen_NJP_Interactioninduced_Currentreversals_2012,Schanz_PRE_Directed_Chaotic_2005,Schanz_PRL_Classical_Quantum_2001}, recent experiments have significantly progressed the realization of highly controllable two dimensional (2D) setups using ac-driven optical lattices \cite{Cubero_PRE_Control_Transport_2012,Denisov_PRL_Vortex_Translational_2008,Lebedev_PRA_Twodimensional_Rocking_2009,Renzoni__Driven_Ratchets_2009} and holographic optical tweezers \cite{Arzola_PRL_Omnidirectional_Transport_2017}. Due to such widespread applications of directed particle transport, the different mechanisms to control the transport have been a topic of ongoing research. One such mechanism is `current reversal' where the direction of the particle transport can be reversed by suitably changing one or more system parameters \cite{Arzola_PRL_Experimental_Control_2011,Cubero_PRE_Control_Transport_2012,Dandogbessi_PS_Controlling_Current_2015,Schreier_EL_Giant_Enhancement_1998,Spiechowicz_CIJNS_SQUID_Ratchet_2019,Wickenbrock_PRE_Current_Reversals_2011,Dinis_PRE_Nonsinusoidal_Current_2015,Zeng_C_Multiple_Current_2012,Kostur_PRE_Multiple_Current_2001,Chen_CS&F_Current_Reversal_2017,Rana_PRE_Current_Reversal_2018}. Indeed, most of the existing schemes to generate current reversals focus on reverting the direction of asymptotic particle transport due to a change of system parameter \cite{daSilva_C_Optimal_Ratchet_2019,deSouzaSilva_N_Controlled_Multiple_2006,Marconi_PRL_Rocking_Ratchets_2007,Mateos_PRL_Chaotic_Transport_2000}. Only recently, research has focused on setups where the current reversal occurs dynamically in time either due to a time-dependent switching of system parameters or due to the presence of interactions and dimensional coupling \cite{Arzola_PRL_Experimental_Control_2011,Liebchen_NJP_Interactioninduced_Currentreversals_2012,Mukhopadhyay_PRE_Dimensional_Couplinginduced_2018,Mukhopadhyay_PRE_Freezing_Accelerating_2016}.

Here, we present a scheme to dynamically generate multiple current reversals due to superimposed driven lattices in two dimensions. The setup employs a `background lattice' driven by an external bi-harmonic oscillating driving force, whose underlying potential is separable in terms of the spatial coordinates. This allows directed transport of particles along the direction of the driving force and trapped motion in the orthogonal direction. On superimposing a second lattice in a finite region of space along the direction of transport leads to a reflection behavior and hence generates a current reversal. Subsequently, the superposition of a third identical lattice can reflect the transport direction once again yielding a second reversal of transport. The timescales of the current reversals can be controlled by the spatial locations of the superimposed lattices. The underlying principle behind the current reversals lie in the conversion of the particle dynamics from chaotic to ballistic in the setup involving multiple lattices, a phenomenon which is forbidden in the background lattice alone. Our paper is structured as follows. In section~\ref{setup}, we describe the underlying setup in details and discuss its relevant symmetries followed by the main results in section~\ref{results}. We discuss the cause of the current reversals in terms of the underlying phase space in section~\ref{discussion}. Finally in section~\ref{experiment}, we provide possible a experimental realization of our setup and conclude our findings in section~\ref{conclusion}.

 \begin{figure}
 	\centering
 	\includegraphics[width=1.0\textwidth]{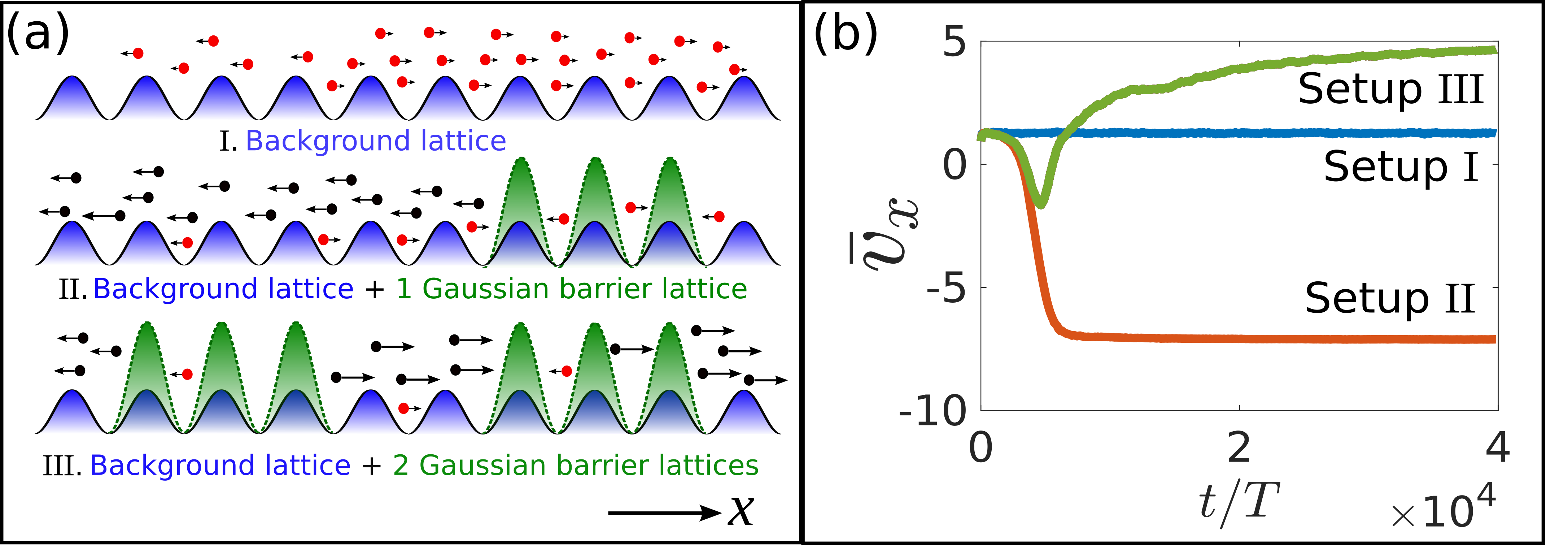}
 	\caption{(\textbf{a}) Schematic representation of a slice of our 2D setup along the $x$-direction and viewed along the $y$-direction. The filled dots denote particles and the colors red and black denote diffusive and ballistic motion respectively. The arrows denote the direction of motion of the particles at asymptotic timescales ($t = t_f$) with the length of the arrow being proportional to the magnitude of the $x$ component of their velocities, i.e. $v_x$. In the presence of only the driven background lattice $V_B$ depicted in blue (setup I, upper panel), most particles exhibit diffusive/chaotic transport towards right, hence the average transport is along the positive $x$-direction. On superimposing a finite lattice of 2D Gaussian barriers $V_{G1}$ (in green), most particles undergo a conversion from diffusive to ballistic motion leading to a reversal of their average transport direction (setup II, middle panel). Their velocities can be reversed once again due to the superposition of a second identical lattice of Gaussian barriers $V_{G2}$, thus leading to a second current reversal. The external driving force is along the $x$-direction. (\textbf{b}) Mean transport velocity of the ensemble along the $x$-direction as a function of time for the three different setups. $U_G({\bf r})=0$ for setup I. For the setup II, $U_G({\bf r})=5$ for $5\times 10^3 <x<10^4$ and vanishes elsewhere whereas for setup III, $U_G({\bf r})=5$ for $5\times 10^3 <|x|<10^4$ and zero elsewhere. Remaining parameters: $U_B=1.0$, $\beta=5$, $d=0.5$.}\label{fig1}
 \end{figure}

\section{Setup, equations of motion and symmetries}\label{setup}
We consider $N$ non-interacting classical particles of mass $m$ in a two dimensional (2D) periodic potential $V({\bf r})=V_B({\bf r}) + V_G({\bf r})$. The separable potential due to the `background lattice' is represented by $V_B({\bf r})= \tilde{V}_B(\cos ^2 \frac {\pi x}{l} +\cos ^2 \frac{\pi y}{l} )$ with potential height $\tilde{V}_B$ and spatial period $l$ in both $x$ and $y$ directions. On top of the lattice $V_B$, we superimpose two finite lattices of 2D Gaussian barriers $V_{G1}$ and $V_{G2}$ localized in different regions which can be described by the potential $V_G({\bf r})=\sum_{m,n=-\infty}^{+\infty} \tilde{U}_G({\bf r}) e^{-\alpha \left( {\bf r} - {\bf r}_{mn}\right)^2}$ with the barriers centered at positions ${\bf r}_{mn}=(ml,nl)$ where $m,n$ $\in\mathbb{Z}$ (see Fig.~\ref{fig1}a). These two lattices also have spatial period $l$ along both $x$ and $y$ directions. The potential height $\tilde{U}_G({\bf r})$ of the barriers depend on their spatial location and $\alpha$ is a measure of the widths of the barriers. In addition, the lattices are driven by an external bi-harmonic periodic driving force $\mathbf{f_D}(t)=a(\cos \omega t + 0.5 \cos 2\omega t,0)$ along the $x$-direction with driving amplitude $a$ and frequency $\omega$. This force is spatially independent. Introducing dimensionless variables $x'=\frac{x}{l}$, $y'=\frac{y}{l}$ and $t'=\omega t$ and dropping the primes for simplicity, the equation of motion for a single particle at position  ${\bf r}=(x,y)$ with velocity ${\bf \dot{r}}=(\dot{x},\dot{y})$ reads
\begin{eqnarray}
\ddot{\bf r} &=& \mathbf{F_B}({\bf r}) + \mathbf{F_G}({\bf r}) + \mathbf{F_D}(t) \nonumber \\
 &=&  U_B \pi (\sin 2\pi x, \sin 2\pi y) +2 U_G({\bf r})\beta \sum_{m,n=-\infty}^{+\infty}   \left( {\bf r} - {\bf R}_{mn} \right) e^{-\beta({\bf r} - {\bf R}_{mn})^2} +  d(\cos t + 0.5 \cos 2t,0) \label{eqm1}
\end{eqnarray}
where $\mathbf{F_B}({\bf r}), \mathbf{F_G}({\bf r})$ and $\mathbf{F_D}(t)$ denote the forces due to the background lattice, superimposed lattices of Gaussian barriers and external driving respectively. The system is described by the four dimensionless parameters: $U_B=\frac{\tilde{V}_B}{ml^2 \omega^2}$ denoting the effective potential height of the lattice $V_B$, $U_G({\bf r})=\frac{\tilde{U}_G({\bf r})}{ml^2 \omega^2}$ denoting the effective potential heights of the Gaussian barriers, $\beta=\alpha l^2$ and the effective driving amplitude $d=\frac{a}{ml\omega^2}$. ${\bf R}_{mn}=(m,n)$ denote the positions of the maxima of the Gaussian barriers which coincides with the positions of the potential maxima of the background lattice $V_B$. In this dimensionless form, the system has a spatial period $L=1$ in both $x$ and $y$ directions and a temporal period $T=2\pi$. 


Our setup breaks the generalized time reversal symmetry $S_{t}$: $t\longrightarrow -t + \tau$, $\mathbf{r}\longrightarrow \mathbf{r} + \pmb{\delta}$ (for arbitrary constant translations $\pmb{\delta}$ and $\tau$ of space and time respectively) and the generalized parity symmetry $P_x$: $x\longrightarrow -x + \delta$, $t\longrightarrow t + \tau$ in the $x$-direction. As a result, directed transport of a particle ensemble is expected along the $x$-direction \cite{Denisov_PR_Tunable_Transport_2014}. Since the setup preserves the generalized parity symmetry along the $y$-direction: $P_y$: $y\longrightarrow -y + \delta$, $t\longrightarrow t + \tau$, directed transport is not possible along this direction. Throughout the following discussions, by `transport' we would always refer to the directed transport along the $x$-direction.

\section{Results}\label{results}
In order to explore the transport properties of our setup, we initialize $N=10^4$ particles within a square region $x,y \in [-5,5] \times [-5,5]$ with small random velocities $v_x,v_y \in [-0.1,0.1] \times [-0.1,0.1]$. The initial velocities of the particles are chosen randomly within the low velocity regime such that their initial kinetic energies are small compared to the potential heights of the lattices. Subsequently we time evolve our ensemble up to time $t = t_f= 4\times 10^4 T$ by numerical integration of Eq.~\ref{eqm1} using a Runge-Kutta Dormand Prince integrator \cite{Dormand_JCAM_Family_Embedded_1980}. We now discuss the transport properties of our setup characterized by the average velocity $\bar{v}_x$ of the particle ensemble along the $x$-direction.

In the presence of only the background lattice $V_B$ (setup I in Fig.~\ref{fig1}a), the particles exhibit directed transport along the positive $x$-direction with an asymptotic transport velocity $\bar{v}_x \simeq 1.3$ (Fig.~\ref{fig1}b). In the setup II we consider a spatially localized lattice of Gaussian barriers $V_{G1}$ superimposed on the lattice $V_B$ between $x=5\times 10^3$ and $x=10^4$ (Fig.~\ref{fig1}a), such that $U_G({\bf r})=5$ for $5\times 10^3 <x<10^4$ and $U_G({\bf r})=0$ everywhere else. In this case, we observe an initial directed transport along the positive $x$-direction with $\bar{v}_x>0$ (Fig.~\ref{fig1}b). However, the transport velocity starts to decelerate and at $t\simeq 3.1\times 10^3 T$, the transport completely vanishes. Thereafter, the ensemble is transported along the negative $x$-direction with $\bar{v}_x<0$ and the transport velocity finally saturates to $\bar{v}_x\simeq -7.1$. Hence, a superimposed spatially localized lattice of Gaussian barriers can trigger a current reversal with the reversal timescale in this case given by $t_{r1}=3.1\times 10^3 T$, i.e. when $\bar{v}_x$ changes its sign.

In the third setup (setup III), we consider a second identical lattice of Gaussian barriers $V_{G2}$ superimposed on the lattice $V_B$ between $x=-5\times 10^3$ and $x=-10^4$ (Fig.~\ref{fig1}a), such that now $U_G({\bf r})=5$ for $5\times 10^3 <|x|<10^4$ and $U_G({\bf r})=0$ elsewhere. Up to $t\simeq 4.6 \times 10^3T$, the transport velocity exhibits a similar behavior as that observed in setup II (Fig.~\ref{fig1}b). Thereafter, instead of asymptotically attaining a negative value, the velocity increases steadily, exhibiting a second current reversal at $t=t_{r2}=5.6 \times 10^3 T$ before finally attaining a constant value $\bar{v}_x\simeq 4.7$. The timescales of current reversal can be controlled by the locations of the lattices $V_{G1}$ and $V_{G2}$. Overall, this demonstrates a controllable scheme to generate multiple reversals of directed particle transport by superimposing spatially localized lattices of 2D Gaussian barriers over a background lattice.

\section{Discussion}\label{discussion}
The mechanism behind such controllable multiple current reversals in our setup crucially depends on the structure of the phase space underlying the system. Since the particles are non-interacting and can move along both $x$ and $y$ directions, the single particle phase space in our externally driven lattice setup is five-dimensional (5D); characterized by $(x,p_x,y,p_y,t)$. However in the absence of the lattices $V_{G1}$ and $V_{G2}$, the particle dynamics along $x$ and $y$ directions can be completely decoupled. Hence the dynamics of the particles in the background lattice $V_B$ driven along the $x$-direction can be described in terms of a three-dimensional (3D) phase space characterized by $(x,p_x,t)$ along $x$ and a 2D phase space characterized by $(y,p_y)$ along $y$ direction. Since we are only interested in the transport along the $x$-direction, we would henceforth only refer to the 3D phase space along the $x$-direction in the course of our discussion.

\begin{figure}
	\centering
	\includegraphics[width=1.0\textwidth]{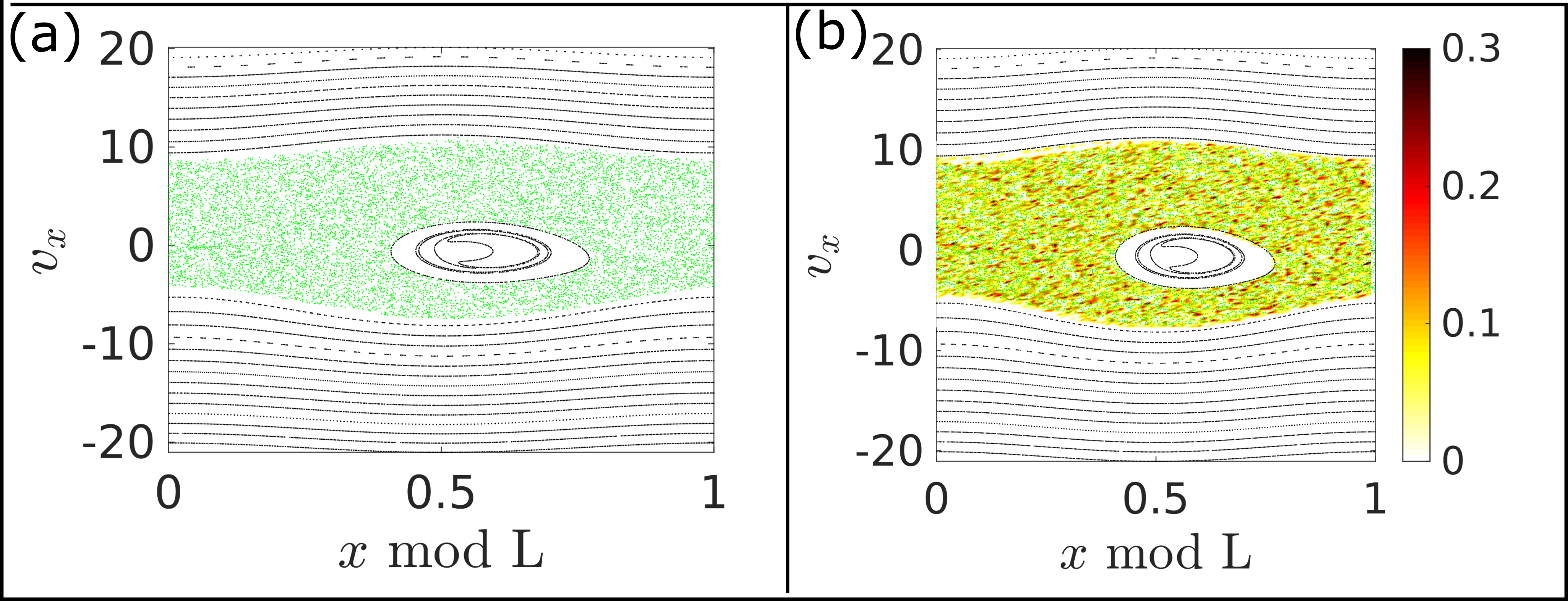}
	\caption{(\textbf{a}) The stroboscopic 2D Poincar\'{e} surface of sections (PSOS) in the $(x,v_x)$ plane corresponding to the driven background lattice $V_B$. The regular islands and invariant curves (in black) denote trapped oscillations and ballistic motion respectively. The chaotic sea (in green) denotes diffusive motion. (\textbf{b}) The asymptotic particle distribution as a function of position $x$ mod $L$ and $v_x$ (in colormap) of all the $N=10^4$ particles propagating in the setup I superimposed on the PSOS shown in Fig.~\ref{fig2}a. The parameters are the same as in Fig.~\ref{fig1}.}\label{fig2}
\end{figure}

\subsection{Directed transport in background lattice}
First, we discuss the directed transport of particles in the positive $x$-direction in the presence of only the lattice $V_B$ and the driving force. In order to do so, it is necessary to understand the phase space underlying our setup I by taking stroboscopic snapshots of particle trajectories $x(t), v_x(t)$ at $t=n T (n\in \mathbb{N})$ with each particle having different initial conditions. This leads to the 2D Poincar\'{e} surface of sections (PSOS): $\{x(nT)\ mod\ L,\ v_x(nT)\}$, which provide a representative overview of the structure of the complete 3D phase space (Fig.~\ref{fig2}a). Due to the broken $P_x$ and $S_t$ symmetries, the PSOS do not possess any reflection symmetry about $v_x=0$. The PSOS is characterized by a single chaotic manifold or `chaotic sea' bounded by the two first invariant spanning curves (FISC) at $v_x\simeq 10$ and $v_x\simeq -6$. The chaotic sea correspond to trajectories undergoing diffusive motion through the lattice. The large regular island embedded in the chaotic sea denotes trapped particles oscillating near the potential minima of the lattice. The particles with speed $|v_x|$ higher than the speed of the respective FISC at positive and negative velocities correspond to ballistic unidirectional motion through the lattice along positive or negative $x$-directions. 

The low energy initial coordinates of our particle ensemble correspond to trajectories in the chaotic sea. Hence in the course of their time evolution, they ergodically populate the entire chaotic sea. This can be observed from (Fig.~\ref{fig2}b), where we project the snapshot of the ensemble population distribution as a function of the particle coordinates $(x,v_x)$ at time $t=t_f$ onto the PSOS. This leads to a converged value of the ensemble velocity which is equal to the transport velocity of the chaotic manifold \cite{Schanz_PRE_Directed_Chaotic_2005}. Physically this signifies that these particles undergo diffusive motion through the lattice which is however asymmetric about $v_x=0$ due to the broken symmetries. Hence, the asymptotic average velocity of the ensemble is non-zero and the particles exhibit directed transport along $x$-direction with $\bar{v}_x\simeq 1.3$ as observed in Fig.~\ref{fig1}b.

\subsection{First current reversal}
Next, we discuss why the transport velocity is reversed due to the superposition of a localized lattice of 2D Gaussian barriers $V_{G1}$ on the background lattice $V_B$ as in the setup II. Here, the particle dynamics is governed by the 2D PSOS (Fig.~\ref{fig2}a) in the region where only the lattice $V_B$ is present, but by the full 5D phase space in the region $5\times 10^3 <x<10^4$ due to the presence of both the lattices $V_B$ and $V_{G1}$. Although this 5D phase space can not be straightforwardly visualized, it turns out that the cause of current reversal can be explained solely on the basis of the ensemble population in the 2D PSOS in Fig.~\ref{fig2}a. Since the ensemble is initialized near the origin $(0,0)$, the particles initially experience the spatial potential only due to the lattice $V_B$ and hence their initial dynamics is exactly the same as described for setup I in the previous subsection. As a result the initial transport velocity is $\bar{v}_x\simeq 1.3$.

\begin{figure}
	\centering
	\includegraphics[width=1.0\textwidth]{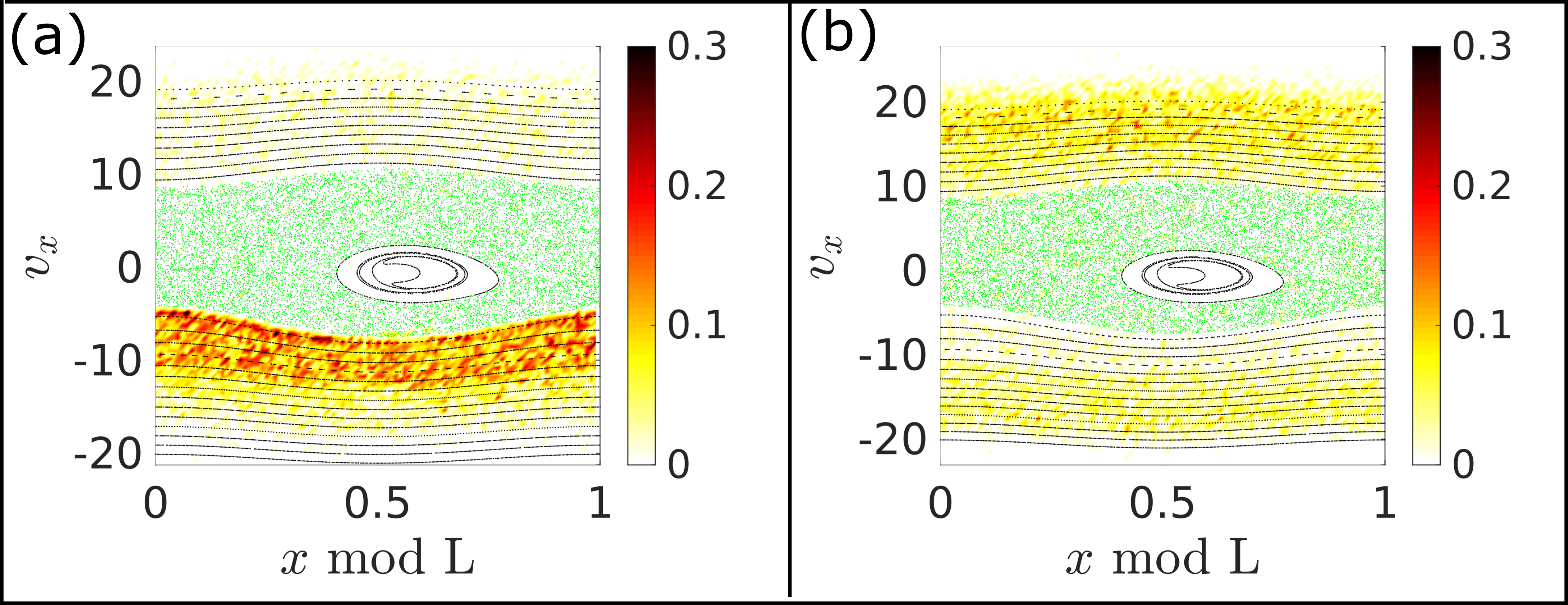}
	\caption{The asymptotic particle distribution as a function of position $x$ mod $L$ and $v_x$ (in colormap) of all the $N=10^4$ particles propagating in the (\textbf{a}) setup II and (\textbf{b}) setup III projected on to the PSOS shown in Fig.~\ref{fig2}a. The parameters are the same as in Fig.~\ref{fig1}.}\label{fig3}
\end{figure}

Since the transport velocity is positive, the particles reach $x=5\times 10^3$ in the course of time where they encounter the lattice $V_{G1}$ in addition to $V_B$. In the region $5\times 10^3 <x<10^4$, since the particle dynamics is no longer governed by the 2D PSOS, the particles are now no longer confined to the central chaotic sea and can access higher velocities beyond the FISC. In fact, the higher potential heights of the Gaussian barriers ensure that most of the particles perform chaotic diffusive motion even at higher velocities corresponding to the full 5D phase space of our setup. This leads to an interesting conversion process between diffusive and ballistic motion of the particles at the left edge of the lattice $V_{G1}$, i.e. at $x=5\times 10^3$, which is the key mechanism behind the current reversal. A diffusive particle close to the left edge but with $x>5\times 10^3$ can cross this edge in the course of time back to $x<5\times 10^3$ with $v_x<0$. However, its velocity $v_x$ immediately after crossing back can be either $\gtrsim -6$ in which case it lies in the chaotic sea performing diffusive motion or $\lesssim -6$ which means it moves away ballistically from the lattice $V_{G1}$ towards the negative $x$-direction. For the particles with $v_x\lesssim -6$, such a conversion from diffusive to ballistic motion ensures that they perform unidirectional ballistic flights towards the negative $x$-direction, thus attaining a permanent negative velocity. On the other hand, since the particles with $v_x\gtrsim -6$ perform diffusive motion they can again enter the region $5\times 10^3 <x<10^4$ in course of time. They would then undergo the same conversion mechanism until all the particles undergo the conversion from diffusive to ballistic motion with $v_x\lesssim -6$. This can be observed in Fig.~\ref{fig3}a where we plot the asymptotic distribution of the ensemble over the 2D PSOS. Most of the particles are located on the invariant curves $v_x\lesssim -6$ moving ballistically in the negative $x$-direction. This results in the asymptotic ensemble transport velocity $\bar{v}_x\simeq -7.1$. The dynamical change in the transport direction leads to the current reversal at $t_{r1}=3.1\times 10^3 T$.

\subsection{Second current reversal}
We now discuss why the transport can be reversed once again by superimposing a second identical lattice of Gaussian barriers $V_{G2}$ between $x=-5\times 10^3$ and $x=-10^4$ as in the setup III. Initially since the particles are initialized near the origin $(0,0)$, the ensemble drifts towards the positive $x$-direction and exhibit the same dynamics as in the setup II. As a result, the transport velocity is initially positive till the first current reversal occurs at $t_{r1}=3.1\times 10^3 T$ and then continues to be negative until $t\simeq 4.6 \times 10^3T$. Thereafter, the particle dynamics undergo another conversion process due to which the transport velocity is reversed again.

Unlike the situation in setup II, the particles in the setup III moving ballistically with negative $v_x$ after $t=t_{r1}$ can not keep moving through the lattice $V_B$ for all longer timescales. Instead at some point, they interact with the lattice $V_{G2}$ in the region $-10^4<x<-5\times 10^3$. Due to the high kinetic energy of the particles (since $|v_x|\gtrsim 6$), some of them can pass through the region and continue their ballistic flights for longer timescales through the lattice $V_B$. This can be seen from the asymptotic ensemble distribution projected on to the 2D PSOS in Fig.~\ref{fig3}b, showing that even at $t=t_f$ a considerable fraction of the ensemble moves with $v_x\lesssim -6$.

However, once a particle enters the region $-10^4<x<-5\times 10^3$, its dynamics is no longer confined to the region $v_x\lesssim -6$ of the 2D PSOS and can explore the different regions of the 5D phase space. Hence most of the particles attain $v_x>0$ which in turn allow them to cross the right edge of the lattice $V_{G2}$ at $x=-5\times 10^3$ back into the region $-5\times 10^3<x<5\times 10^3$ where only the lattice $V_B$ is present. After crossing to this region, these particles can either belong to the chaotic sea or to the invariant spanning curves with velocity higher than the FISC at $v_x\simeq 10$ of the 2D PSOS in Fig.~\ref{fig2}a. The particles with $v_x\gtrsim 10$ perform unidirectional ballistic flights in the positive $x$-direction. Due to their significantly higher kinetic energy, these particles are not `reflected' further by the lattice $V_{G1}$ in the region $5\times 10^3 <x<10^4$; instead, after crossing this region, they continue moving ballistically through the lattice $V_B$ with $v_x\gtrsim 10$. This can be observed from the significant distribution of particles with $v_x\gtrsim 10$ in Fig.~\ref{fig3}b at $t=t_f$. As the velocities of more and more particles undergo the conversion from $v_x\lesssim -6$ to $v_x\gtrsim 10$, the transport velocity increases steadily after $t\simeq 4.6 \times 10^3T$, leading to a second current reversal at $t=t_{r2}=5.6 \times 10^3 T$ (Fig.~\ref{fig1}b). On the other hand, the particles in the chaotic layer would eventually again encounter the lattices $V_{G1}$ or $V_{G2}$ so that their chaotic dynamics is eventually converted to ballistic motion either with $v_x\gtrsim 10$ or with $v_x\lesssim -6$. Due to the overall higher number of particles moving asymptotically with $v_x\gtrsim 10$ compared to those with $v_x\lesssim -6$ (see Fig.~\ref{fig3}b), the asymptotic transport velocity is $\bar{v}_x\simeq 4.7$.

\section{Experimental realization}\label{experiment}
Our scheme of multiple current reversal can be experimentally realized using cold atoms or colloids with optical lattices \cite{Brown_PRA_Ratchet_Effect_2008,Renzoni__Driven_Ratchets_2009,Arzola_PRL_Experimental_Control_2011} and lattices designed using holographic trapping techniques \cite{Barredo_S_Atombyatom_Assembler_2016,Kim_NC_Situ_Singleatom_2016,Nogrette_PRX_SingleAtom_Trapping_2014,Stuart_NJP_Singleatom_Trapping_2018,Arzola_PRL_Omnidirectional_Transport_2017}. The background lattice can be formed by 2D optical lattices where the periodic potential is generated by counterpropagating laser beams of perpendicular polarization. The spatially localized lattices of 2D Gaussian barriers can be obtained by reflecting a linearly polarized laser beam onto a spatial light modulator (SLM) displaying a computer generated hologram. The external driving force can be realized using a piezo-modulator \cite{Arzola_PRL_Omnidirectional_Transport_2017}.

Translating our parameters to experimentally relevant quantities for an optical lattice setup with cold rubidium (Rb$^{87}$) atoms and $780\ nm$ lasers, we obtain the lattice height $\tilde{V}_B \sim 5 E_r$, the width $\frac{1}{\sqrt{\alpha}}\sim 350\ nm$, the driving frequency $\omega \sim \omega_r$ and the driving amplitude $a \sim 0.003\ E_r/nm$, where $E_r$ and $\omega_r$ are the recoil energy and recoil frequency of the atom respectively. The timescales of the current reversals can be controlled by the spatial locations of the two lattices of Gaussian barriers. Further away the lattices are from the origin, i.e. near the initial location of the ensemble, the larger would be the reversal timescales. In contrast to Brownian ratchets, our mechanism does not depend on noise and operates in the purely Hamiltonian regime. The effect of weak noise typical for ratchet experiments with cold atoms and underdamped colloids \cite{Cubero_PRE_Current_Reversals_2010,Hanggi_RMP_Artificial_Brownian_2009} represent minor fluctuations of the average velocity of the ensemble and this does not affect the functionality of the current reversal mechanism. Interaction and disorder have been shown to enhance accumulation of particles within the regular regions of the phase space \cite{Liebchen_NJP_Interaction_Induced_2015,Liebchen_NJP_Interactioninduced_Currentreversals_2012,Wulf_PRL_Disorder_Induced_2014}, which would aid the conversion of chaotic to ballistic dynamics of particles. This would possibly decrease the reversal timescales.

\section{Brief conclusions}\label{conclusion}
We provide a scheme to realize time dependent multiple reversals of directed transport in a two dimensional driven lattice setup by superimposing `spatially localized lattices' on top of a `global background lattice'. In contrast to most other current reversal schemes, the reversal of transport here occurs dynamically and the timescales of reversal can be controlled by controlling the spatial location of the localized lattices. The scheme is generic in the sense that the only requirement is a mixed phase space corresponding to the underlying background lattice and hence can be applied to a variety of physical systems, for e.g, cold atoms and colloids.





\vspace{6pt} 



\authorcontributions{conceptualization, A.K.M. and P.S.; methodology, A.K.M.; software, A.K.M.; validation, A.K.M.; formal analysis, A.K.M; investigation, A.K.M; resources, P.S.; data curation, A.K.M.; writing--original draft preparation, A.K.M.; writing--review and editing, A.K.M. and P.S.; visualization, A.K.M.; supervision, P.S.; project administration, P.S.; funding acquisition, A.K.M. and P.S.}

\funding{A.K.M acknowledges a doctoral research grant (Funding ID: 57129429) by the Deutscher Akademischer Austauschdienst (DAAD).}

\acknowledgments{The authors thank B. Liebchen and T. Wulf for insightful discussions.}

\conflictsofinterest{The authors declare no conflict of interest. The funders had no role in the design of the study; in the collection, analyses, or interpretation of data; in the writing of the manuscript, or in the decision to publish the results.} 

%

%

\reftitle{References}


\externalbibliography{yes}
\bibliography{mybib,revtex_custom}





\end{document}